\documentclass[a4paper,fleqn,usenatbib]{mnras}

\usepackage{hyperref}

\usepackage{newtxtext,newtxmath}

\usepackage[T1]{fontenc}
\usepackage{ae,aecompl}

\usepackage{graphicx}	
\usepackage{graphicx}	
\usepackage[shortlabels]{enumitem}      
\usepackage{svg}    
\usepackage{booktabs}   
\usepackage{algorithm}  
\usepackage[noend]{algpseudocode}  
\usepackage{bm}
\makeatletter
\def\BState{\State\hskip-\ALG@thistlm}
\makeatother

\usepackage[export]{adjustbox}

\usepackage{amsmath}

\usepackage{subcaption}

\definecolor{orange}{rgb}{0.8,0.35,0}
\definecolor{purple}{rgb}{0.3,0.15,0.55}

\graphicspath{{Figures}}
\DeclareGraphicsExtensions{.png,.pdf,.eps}

\defcitealias{Semelin2017}{21SSD}
\defcitealias{Karras2017}{PGGAN}

\title[Local environment's effect on WL statistics]{The density distributions of cosmic structures: impact of the local environment on weak-lensing convergence}

\author[Ema S. A., \textit{et al.}]{
Sonia Akter Ema$^{a}$\thanks{E-mail: sema0128@uni.sydney.edu.au}, Md Rasel Hossen$^{a}$, Krzysztof Bolejko$^{b}$ and Geraint F. Lewis$^{a}$
\\
$^{a}$Sydney Institute for Astronomy, School of Physics, A28, The University of Sydney, NSW 2006, Australia \\
$^{b}$School of Natural Sciences, College of Sciences and Engineering, University of Tasmania, Private Bag 37, Hobart TAS 7001, Australia }

\date{Accepted XXX. Received YYY; in original form ZZZ}

\pubyear{2021}

\begin{document}
\label{firstpage}
\pagerange{\pageref{firstpage}--\pageref{lastpage}}
\maketitle

\begin{abstract}
Whilst the underlying assumption of the Friedman-Lemaître-Robertson-Walker (FLRW) cosmological model is that matter is homogeneously distributed throughout the universe, gravitational influences over the life of the universe have resulted in mass clustered on a range of scales. Hence we expect that, in our inhomogeneous universe, the view of an observer will be influenced by the location and local environment. Here we analyse the one-point probability distribution functions and angular power spectra of weak-lensing (WL) convergence and magnification numerically to investigate the influence of our local environment on WL statistics in relativistic $N$-body simulations. To achieve this, we numerically solve the null geodesic equations which describe the propagation of light bundles backwards in time from today, and develop a ray-tracing algorithm, and from these calculate various WL properties. Our findings demonstrate how cosmological observations of large-scale structure through WL can be impacted by the locality of the observer. We also calculate the constraints on the cosmological parameters as a function of redshift from the theoretical and numerical study of the angular power spectrum of WL convergence. This study concludes the minimal redshift for the constraint on the parameter $\Omega_m$ ($H_0$) is $z \sim 0.2$ $(z \sim 0.6 )$ beyond which the local environment's effect is negligible and the data from WL surveys are more meaningful above that redshift. The outcomes of this study will have direct consequences for future  surveys, where percent-level-precision is necessary.
\end{abstract}


\section{Introduction}
 Upcoming galaxy surveys, such as the  Large Synoptic Survey Telescope \citep{LSST2012}, Dark Energy Spectroscopic Instrument \citep[DESI;][]{Aghamousa2016}, Evolutionary Map of the universe \citep[EMU;][]{Norris2011}, Euclid \citep{Amendola2018}, and others \citep{Santos2015, Walcher2019} will  map the  universe over large-scales and to significant depth. These surveys will measure various cosmological properties (viz. redshift, brightness, shape, and sizes of galaxies, etc.) with unprecedented precision and thus quantitative understating of the influence of local cosmological environment on cosmological observation is essential. 

\par Gravitational lensing is a unique tool to obtain information about   the distribution of matter around the lensing objects. In an inhomogeneous universe, all under- and over-densities of matter act as gravitational lenses and each light ray from distant galaxies is sheared differently, leading to an observer-dependent view based upon their location within the large-scale structure (LSS) of the universe. 
The study of gravitational lensing magnification, shear, convergence, and other cosmological properties by investigating WL is an established approach in astronomy \citep{Kaiser1992, Kaiser2000,  Maoli2001, Brown2002, Barber&Taylor2003, Killedar2012, Takahashi2012, Takahashi2017}. However, a relatively small number of studies have focused upon numerically solving the null geodesic equations in order to determine the light propagation effects in inhomogeneous cosmologies as light traverses significant distances \citep{2011MNRAS.412.1937B,2011JCAP...02..025B, Killedar2012, 2012JCAP...05..003B, Yamauchi2013, Thomas2015a}. Previously, the distance redshift relation was computed through the investigation based on a simplified model for inhomogeneous universe \citep{Watanabe&Tomita1990}. The WL was numerically studied by \cite{Couchman1999}, where they constructed a 3D ray-tracing code and discussed the difference between 2D shear and 3D shear. An approach to study WL by voids was presented by \citet{Barreira2015}. Their outcomes suggest that the observation of WL signals associated with under-dense regions is a promising tool to constrain the law of gravitation on large-scales. \citet{Reischke2019} have theoretically investigated how the cosmological observables depend on the locality of an observer. After critically examining the analysis of \citet{Reischke2019}, \citet{Hall2020} has concluded that there is no evidence of significant bias depending on the observer's position at different regions on the LSS. 

\par Numerical simulations play a vital role to study the evolution of the LSS of the universe. Typical approaches have employed Newtonian gravity-based $N$-body simulations \citep{Dyer&Roedar1974, Futamase&Sasaki1989, Babul&Lee1991, Casarini2012, Borzyszkowski2015, Renneby2018} to study the nonlinear structure formation of the galaxies. These simulations consider that the gravitational fields are weak and they originate from non-relativistic matter, it's important to add relativistic effects in $N$-body simulation to understand properly the "dark" side of the universe. In this paper we use numerical simulations based on \texttt{gevolution} \citep{Adamek2016, Adamek2016b}, which use relativistic $N$-body approaches based on the weak field approximation. Considering the relativistic properties, this can compute all six metric degrees of freedom in Poisson gauge. Very recently, \cite{Lepori2020} have investigated the WL observables using this \texttt{gevolution} code. They have studied the relativistic ellipticity, convergence, image rotation, and two-point correlation functions as a function of cosmological distance by implementing a completely different methodology from us, they did not consider the local environment's dependency on LSS there. In our present work, we have used this relativistic $N$-body code to generate the weak perturbations.

\par This work contains the results from our relativistic $N$-body simulations to study the WL statistics. We clone \texttt{gevolution} from the github repository\footnote{\url{https://github.com/gevolution-code/gevolution-1.2}} and generate snapshots for gravitational potentials and particle's position at different redshifts setting the simulation parameters properly. After that we construct a new ray-tracing algorithm by integrating null geodesic equations from today towards Big Bang and then run our own algorithm to study the WL properties. The background of our simulation is the standard $\Lambda$CDM cosmological model. The code computes the convergence for various line of sights. From these we compute the maps of the convergence, shear, and magnification. The maps are then used to infer WL probability distribution functions (PDFs) and angular power spectra. These are then examined for the location of the observer and its cosmological environment.

\par This paper is structured as follows: Section \ref{sec:RAY-TRACING} contains the description of our ray-tracing algorithm and how we calculate the WL properties. In Section \ref{sec:SIMULATION} we provide the information about our $N$-body simulation and the adopted algorithms to find voids and halos from the simulation in Section \ref{sec: Finding voids and clusters}. The theoretical approach to calculate the WL convergence angular power spectrum is discussed in Section \ref{sec: Theoretically calculate convergence angular power spectra} and in Section \ref{sec: Numerically calculate convergence angular power spectra} we describe how we infer the angular power spectra from the simulations. We devote Section \ref{sec:RESULTS ANALYSIS} to describe the results of this work and finally, we present the conclusions in Section \ref{sec:CONCLUSIONS}.

\section{Ray-tracing}
\label{sec:RAY-TRACING}

\subsection{Weak-lensing theory}

We present here the basic theory behind WL and a number of important results from WL approximation which will require: the WL theory as well as equations to calculate the magnification, cosmic shear, and convergence. 
Convergence is the integrated mass density along the line of sight that comes from the isotropic Ricci focusing, a linear function of the amount of matter, of a beam due to enclosed matter. Due to the WL by the intervening LSS of the universe, images of distant galaxies are experiencing distortion, and this distortion is known as cosmic shear. Convergence and shear contribute together to the area of the source on the sky, whereas the magnification arises from the conservation of the surface brightness. 

If there are two neighbouring geodesics $\mathcal{L}$ and $\mathcal{L^{'}}$ enclosed by a light bundle then assuming the small angle approximation, we can write 
\begin{equation}
\alpha_i(z) = a \, D_{ij}(z) \, \theta_j, 
\label{equation1}
\end{equation}
\noindent where $\alpha$ is the angular diameter distance between $\mathcal{L}$ and $\mathcal{L^{'}}$ at redshift $z$, $a$ is the expansion factor, and $\theta_j$ is the angle between the two neighboring geodesics at the observer's location. $D_{ij}$ provides different information for different universes. For an inhomogeneous universe, $D_{ij}$ indicates the distortion of light bundles that is being produced due to the density distribution. On the other hand, in case of a homogeneous universe, $D_{ij}$ = $D_0(z)$ $\delta^{K}_{ij}$, $D_0$ is the physical distance and $\delta^{K}_{ij}$ is the Kronecker delta. 

Let $\beta$ be the observed position of a source and the true position is $\beta_0$, then the deformation of the image of that source can be represented by the following 2$\times$2 Jacobian matrix
\begin{equation}
\mathbf{A} = \frac{\partial \beta_0^{i}}{\partial \beta_j} \equiv
\begin{bmatrix}
  1 - \kappa - \gamma_1 & 
    - \gamma_2 \\[1ex] 
  - \gamma_2 & 
    1 - \kappa + \gamma_1 \\[1ex]
\end{bmatrix},
\label{Jacobian_matrix}
\end{equation}
\noindent where $\kappa$ is the convergence, $\gamma$ is the shear, and the components of the cosmic shear can be represented by, $\gamma = \sqrt{ \gamma_1^2 + \gamma_2^2 }$ and $\gamma_2$ can be expressed in terms of the second derivatives of the gravitational potential $\psi$ along two orthogonal directions as
\begin{equation}
\gamma_1 = \frac{1}{2} \left( \psi_{, 11} - \psi_{, 22} \right)   
\\ 
\& \\ \gamma_2 = \psi_{,12},
\label{equation_2}
\end{equation}
where the indices after the comma sign indicate partial differentiation. The lens equation in the locally linearised form is
\begin{equation}
    I(\Vec{\alpha}) = I^{(s)}[\Vec{B_0} + \bm{A}\Vec{\alpha_0} \cdot (\Vec{\alpha} - \Vec{\alpha_0})],
\label{lens_equation}
\end{equation}
\noindent where $\Vec{\alpha_0}$ is a point within an image, $\Vec{B_0}$ = $\Vec{B}\Vec{\alpha_0}$ within the source, $\Vec{B}$ is the angular positions of the images of a source, $I_s(\Vec{B})$ is the surface brightness in the source plane, and $I_s(\Vec{\alpha})$ is the observed surface brightness in the lens plane. Equation (\ref{lens_equation}) tells us that the images of circular sources are ellipses and the flux magnification of an image can be calculated from the inverse of the determinant of the Jacobian matrix, \bm{$A$}
\begin{equation}
    \mu = \frac{1}{det \bm{A}} = \frac{1}{(1 - \kappa)^2 - |\gamma|^2}.
\label{mu_equation}
\end{equation}

\subsection{Ray bundle method}
\label{sec:RAY BUNDLE METHOD}

\begin{figure}
  \centering
  \noindent
  \includegraphics[width=\columnwidth]{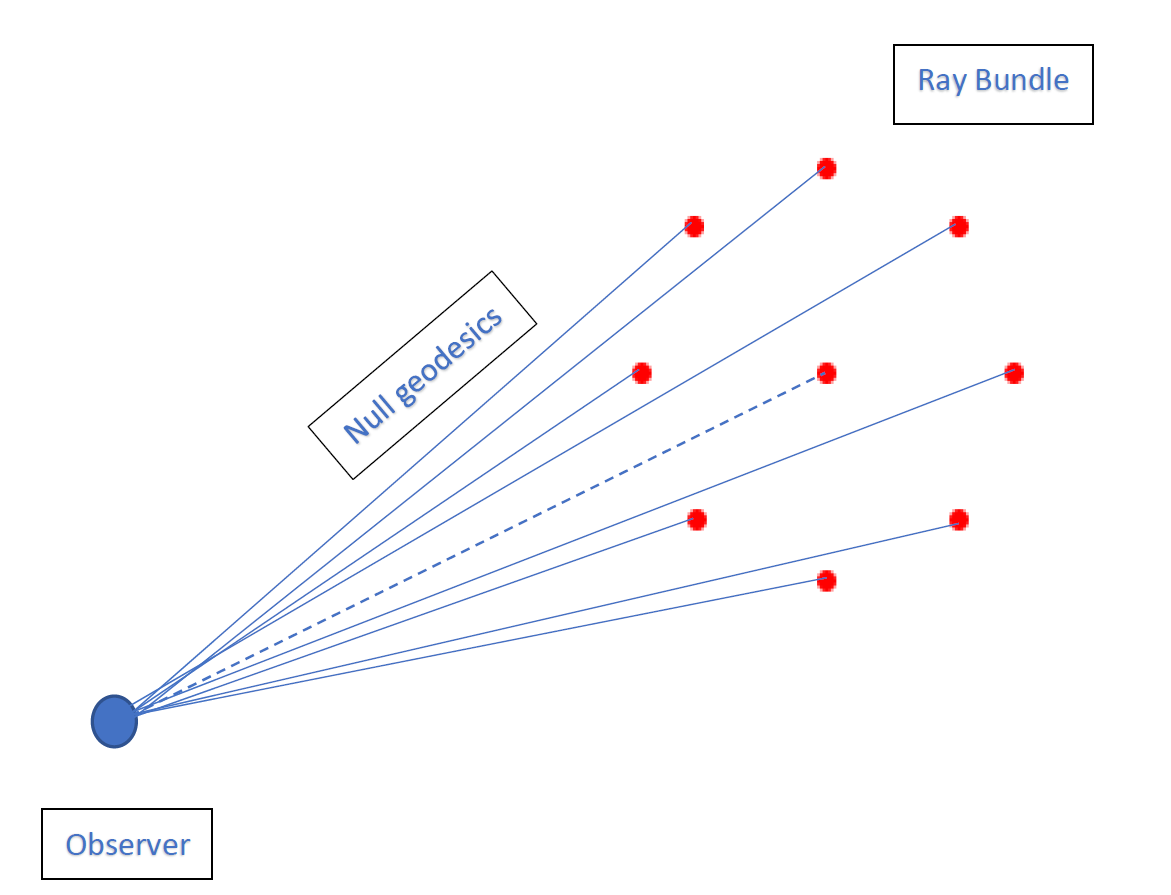}
  \caption{The schematic diagram of RBM. There are nine light rays here that are indicated by nine red points. Each of the light rays represents null geodesic and the central null geodesic here is shown by the blue dashed line.}
   \label{fig:figure2}
\end{figure}

In this paper, we use an alternative technique to the ray-shooting method (RSM) for calculating the magnification, convergence, and cosmic shear in the WL limit, namely, the ray bundle method (RBM). This method was developed by \citet{Fluke1999} \citep[see also][]{Barber2000, Fluke2002, Fluke&Lasky2011, Killedar2012} and it allows to study the WL statistics for cosmological models. The RBM has been developed for situations where the multiple image creation of any source has not taken as a concern. This method works in the similar way to the well-known RSM but they are different from a structural point of view. Instead of modelling individual light rays in the line of sight, the RBM uses the pattern of a ray bundle that contains eight light rays and a central ray. 
Fig \ref{fig:figure2} shows a schematic diagram of RBM, where we show the formation of one bundle employing the RBM. This figure clearly indicates how we have projected light rays as bundles from the observer into space.

\subsection{Magnification}\label{sec:magnification}

Magnification is a geometric consequence of gravitational lensing. The magnification can change the observed size and flux of individual galaxies. There are mainly two techniques to measure magnification, one is from the change in local source counts \citep{Broadhurst1995} and another is from the change of the image sizes when the surface brightness is fixed \citep{Bartelmann&Narayan1995}. When a bundle of light rays pass through a transparent lens, the intensity of the source does not change due to the effect of gravitational lensing, however the cross-sectional area of the light bundle is impacted. According to \cite{Dyer&Roedar1974}, the change in the apparent brightness depends on the change in the solid angle that the image covers and on the presence of a lens, hence
\begin{equation}
\lvert\mu\lvert= \frac{S_{\nu}}{S^{'}_{\nu}},
\end{equation}
\noindent where the flux at a frequency $\nu$ is $S_{\nu} = I_{\nu} \, d\Omega_{obs}$, $\,I_{\nu}$ is the specific intensity and $d \Omega_{obs}$ is the solid angle subtended by the source at the observer's location.

The algorithm implemented in this study projects photons as bundles. At the initial instant each bundle has the same radius. As the bundle propagates its shape and area changes due to the underlying matter distribution along the line of sight. The magnification is then calculated as
\begin{equation}
\mu = \frac{A_{image}}{A_{source}},
\label{mu_RBM}
\end{equation}
\noindent where $\mu$ is the magnification, $A_{image}$ is the area of of image, and $A_{source}$ denotes the area of source in RBM. In a perturbed universe, demagnification ($\mu < 1 $) occurs in under-dense regions and magnification ($\mu > 1 $) takes place in over-dense regions.

\subsection{Cosmic Shear} \label{sec:shear}
The cosmic shear is a powerful probe to study the nature of dark matter as well as the expansion history of the universe. The images of distant galaxies are sheared due to the WL by LSS of the universe \citep{Escude&Jordi1991, Schneider2002}. If a galaxy has an intrinsic and complex source ellipticity $\epsilon^s$ then the cosmic shear can modify this ellipticity as a function of complex reduced shear. The observed ellipticity is \citep{Seitz&Schneider1997}
\begin{equation}
{\epsilon} = \frac{\epsilon^{s} + g}{1 + g^{*}\epsilon^{s}},
\label{shear_galaxy}
\end{equation}
\noindent where $g^{*}$ is the complex conjugate of reduced shear, $g$. In WL limit Eq. (\ref{shear_galaxy}) reduces to  $\epsilon \approx \epsilon^{s} + \gamma$, and $\lvert \epsilon \lvert = g$ when $\lvert \epsilon^{s} \lvert = 0$.

The RBM algorithm implemented in this study allows to infer the ellipticity the bundle (and hence the image ellipticity)
\begin{equation}
\boldsymbol{\gamma} = \frac{c - d}{c + d},
\label{shear_RBM}
\end{equation}
\noindent  where $c$ and $d$ are the semi-minor and -major axes of the bundle, which is obtained by fitting ellipses to the bundle.  


\subsection{Null geodesics and ray-tracing} \label{sec:geodesics}

In this work, we solve the null geodesic equations to observe the distortion of the ray bundles, where the weak gravitational potentials found from the $N$-body code \texttt{gevolution} \citep{Adamek2016} play a vital role. 
Perturbing a spatially flat FLRW metric in the Poisson gauge yields 
\begin{align}  \label{gevo_metric}
ds^2 = a^2(\tau) [- (1+2\psi)d\tau^2 - 2B_ix^id\tau + (1-2\phi)\delta_{ij}dx^idx^j \notag\\
+ h_{ij}dx^idx^j],
\end{align}
 \noindent where $a(\tau)$ is the scale factor, $\tau$ is conformal time, $\psi$ and $\phi$ are scalar perturbations, $B_i$ are vector perturbations, and $h_{ij}$ are tensor perturbations. \texttt{gevolution} implements the same gauge for metric perturbations.

The RBM method implemented in this paper is based on solving null geodesic equations for light rays travelling from observer to source
\begin{equation}
    \frac{d^2x^\alpha}{d\lambda^2} = - \Gamma_{\beta\gamma}^\alpha \frac{dx^\beta}{d\lambda} \frac{dx^\gamma}{d\lambda},
\label{geodesic_equation}    
\end{equation}
\noindent where $\Gamma_{\beta\gamma}^\alpha$ is the Christoffel symbols computed from the metric (\ref{gevo_metric}) with scalar perturbations only. $\lambda$ is the affine parameter and Greek indices can take the values: 0, 1, 2, 3.

In our code we project the light rays as bundles in such a way where each bundle contains $9$ geodesics (including the central ray), the light bundles will start to travel from one box to another box (since in \texttt{gevolution}, they impose the periodic boundary conditions so that when rays leave the box will enter the box again from the opposite side) and after that the shape of the bundles are gradually distorted while the bundles travel through space. At first we save some snapshots at different redshifts by using the $N$-body code \texttt{gevolution}  \citep{Adamek2016}, that contain the information about the scalar potentials in the simulation cube when light travels different cosmological distances. That means the space between an observer and source is divided up into individual regions, then each modelled by a snapshot of a cosmological simulation at an appropriate redshift. Finally, we develop a ray-tracing algorithm in such a way so that it can take the snapshots at different redshifts as inputs as well as the light bundles could travel from one box to another box by maintaining the proper evolution of cosmological properties (e.g., scale factor, redshift, energy).

\begin{figure}
  \centering
  \noindent
  \includegraphics[width=\columnwidth]{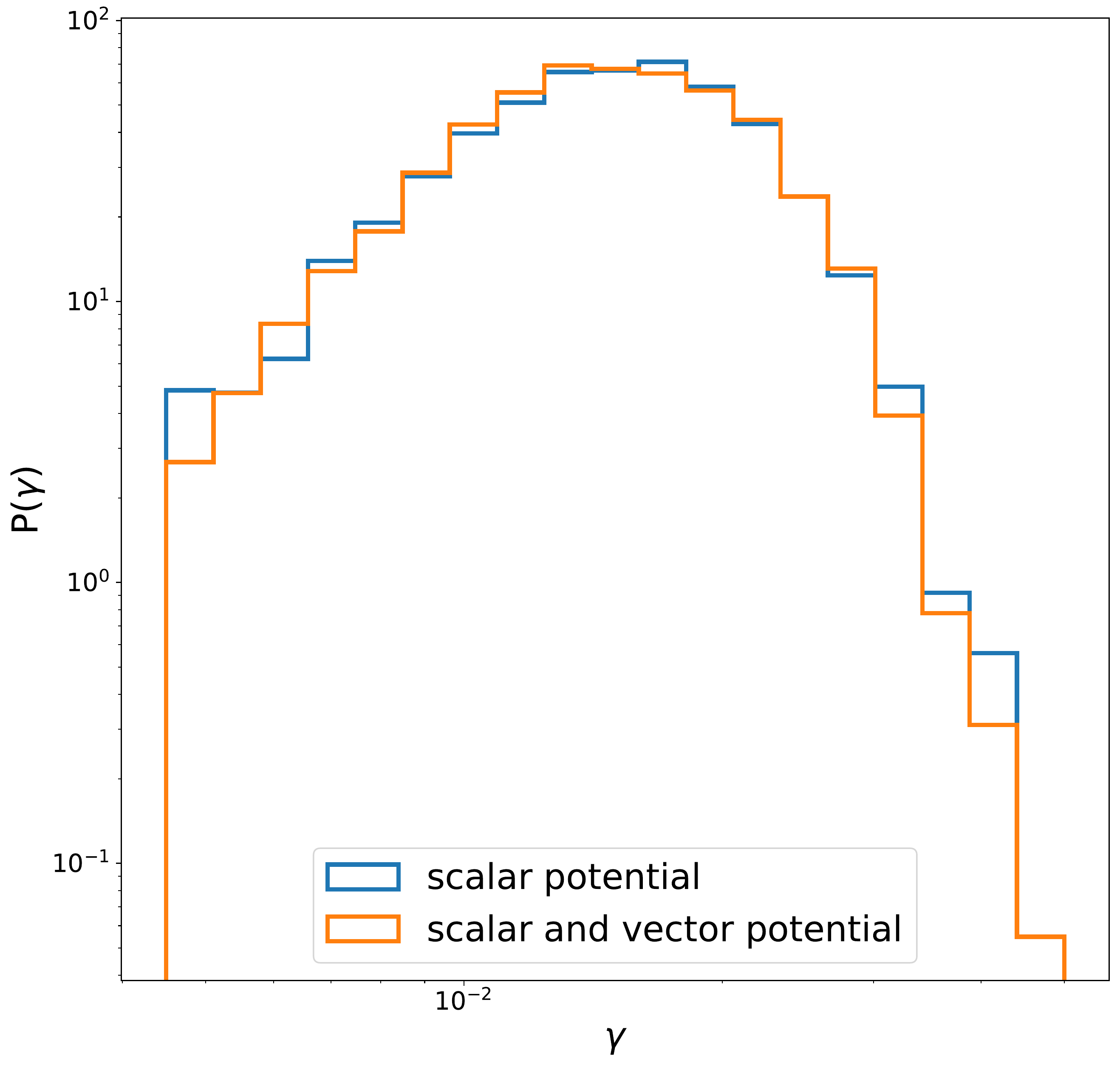}
  \caption{The probability distributions of cosmic shear, $\gamma$, at a redshift of 0.4. The blue line is when the shear probability distribution has been calculated from a metric that depends on scalar potentials, $\psi$ $\&$ $\phi$, only, whereas the orange line represents that the weak field metric contains the components of both scalar and vector potentials, $\psi$, $\phi$, $\&$ $B_i$.}
   \label{fig:scalar_vector_potential}
\end{figure}

The evolution of the metric components is obtained from the $N$-body code \texttt{gevolution}. This is then used to compute the Christoffel symbols and solve the geodesics equations\footnote{We solve 49,152 bundles of geodesics here. This is not a limitation of our ray-tracing algorithm and one can easily solve different number of bundles.}. It is important to mention here that the amplitudes of the vector and tensor perturbations are very small than the scalar potentials, $\psi$ $\&$ $\phi$, \citep{ Lu2009, Thomas2015b, Adamek2016}, so their effects will be negligible in the round-off error when solving geodesic equations at single precision. We perform initial tests to confirm this and then we neglect the effects of the vector and tensor perturbations $B_i$ $\&$ $h_{ij}$ in this work, i.e. the Christoffel symbols here depend on the gradients of the scalar potentials $\psi$ $\&$ $\phi$ only. Fig \ref{fig:scalar_vector_potential} shows the PDFs of cosmic shear, $\gamma$, when the geodesic equations have been solved from the weak field metric that contains only scalar potentials (blue line) and both scalar as well as vector potentials (oragne line). After solving the null geodesic equations from some random realizations, we plot the mean probability distribution curves for both cases here considering the equal binning scale.
As expected, there is no apparent distinction between the obtained PDFs, other than the variations due to low number statistics. Thus, in further studies in this paper we focus on scalar perturbations only and explore how local environment affects the WL observables. In future studies we may include $B_i$ when investigating higher order statistics which could be sensitive to vector perturbations. The radius of the ray bundle has a negligible effect on the results of the WL statistics, and we consider here that all of bundles possess an initial radius of 0.01 [physical units] \citep{Fluke1999}. We set the initial conditions at the present day and trace the bundles back in time up to a comoving distance of 1.5 $\mathrm{Gpc}/h$ (redshift $z \approx 0.62$), where $h$ is the dimensionless Hubble parameter. 
For this we infer a set of WL observables (see Secs. \ref{sec:magnification} and \ref{sec:shear}) for a single observer. We then repeat the procedure for different observers to test the depencey on the local cosmological environment (see Sec. \ref{sec: Finding voids and clusters}).

\begin{figure}
    \centering
    \begin{subfigure}[t]{0.4\textwidth}
        \centering
        \includegraphics[width=1.1\textwidth]{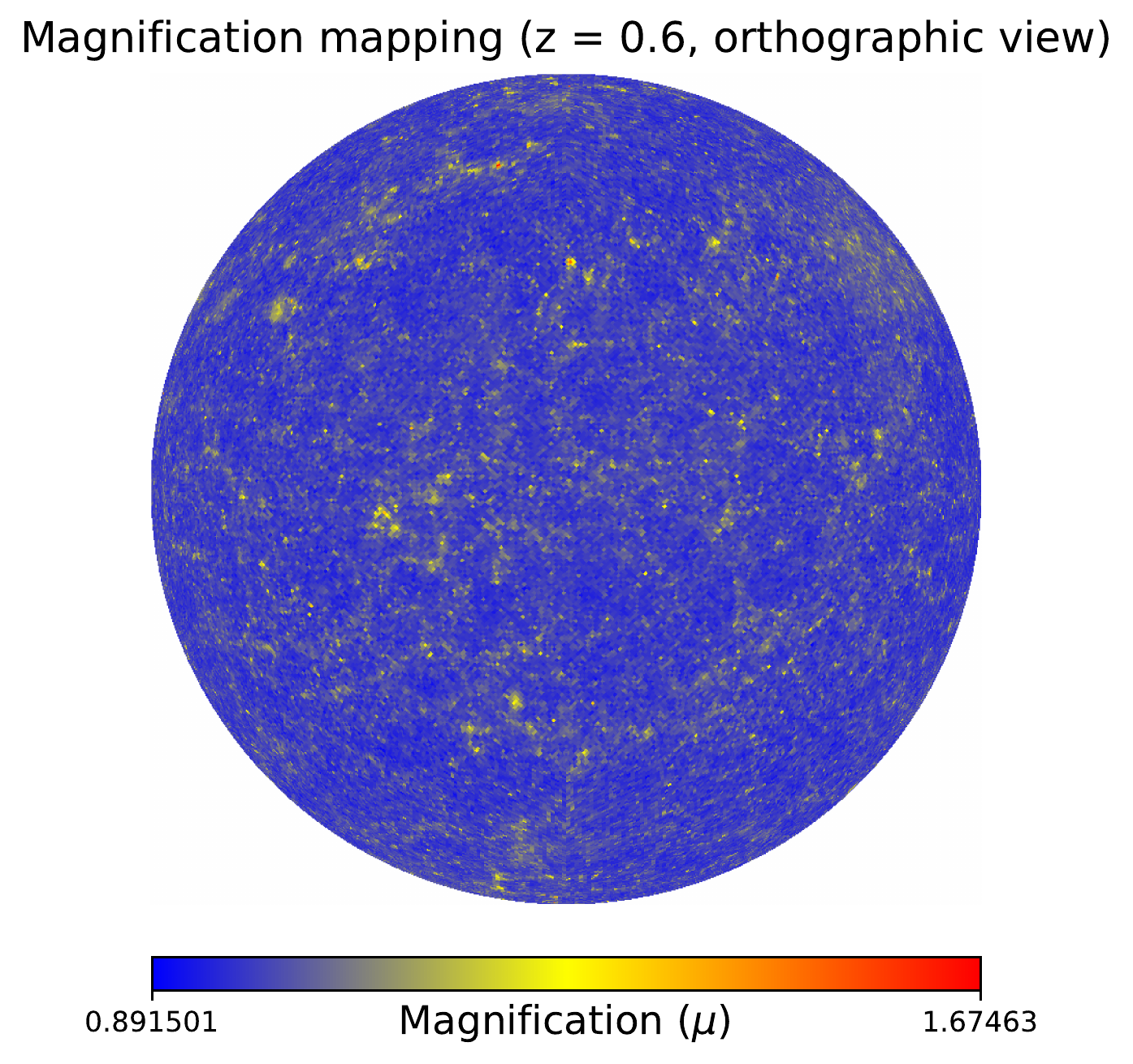}
            \end{subfigure}%
    
    \begin{subfigure}[t]{0.4\textwidth}
        \centering
        \includegraphics[width=1.0\textwidth]{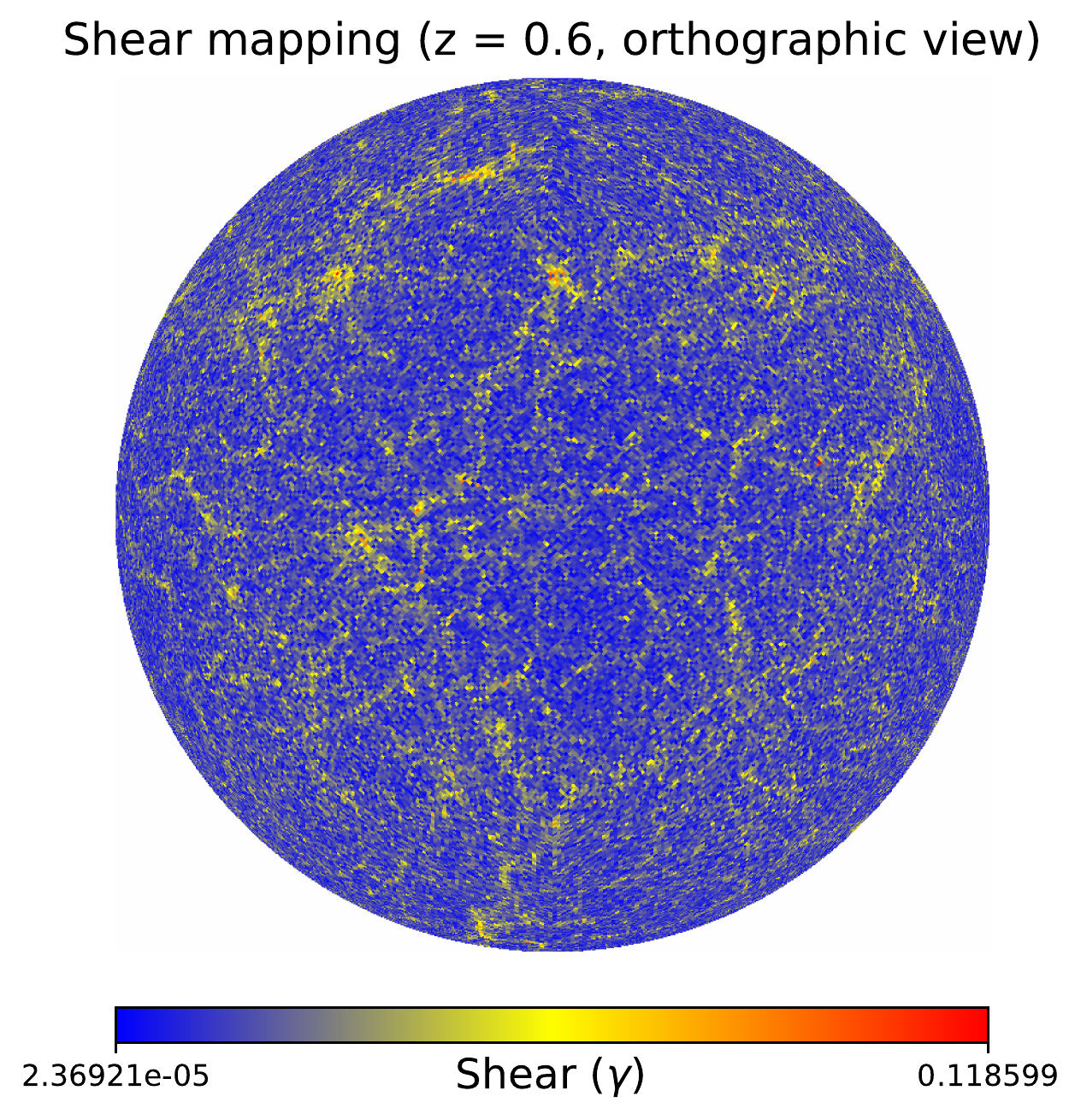}
            \end{subfigure}

    \begin{subfigure}[t]{0.4\textwidth}
        \centering
        \includegraphics[width=1.1\textwidth]{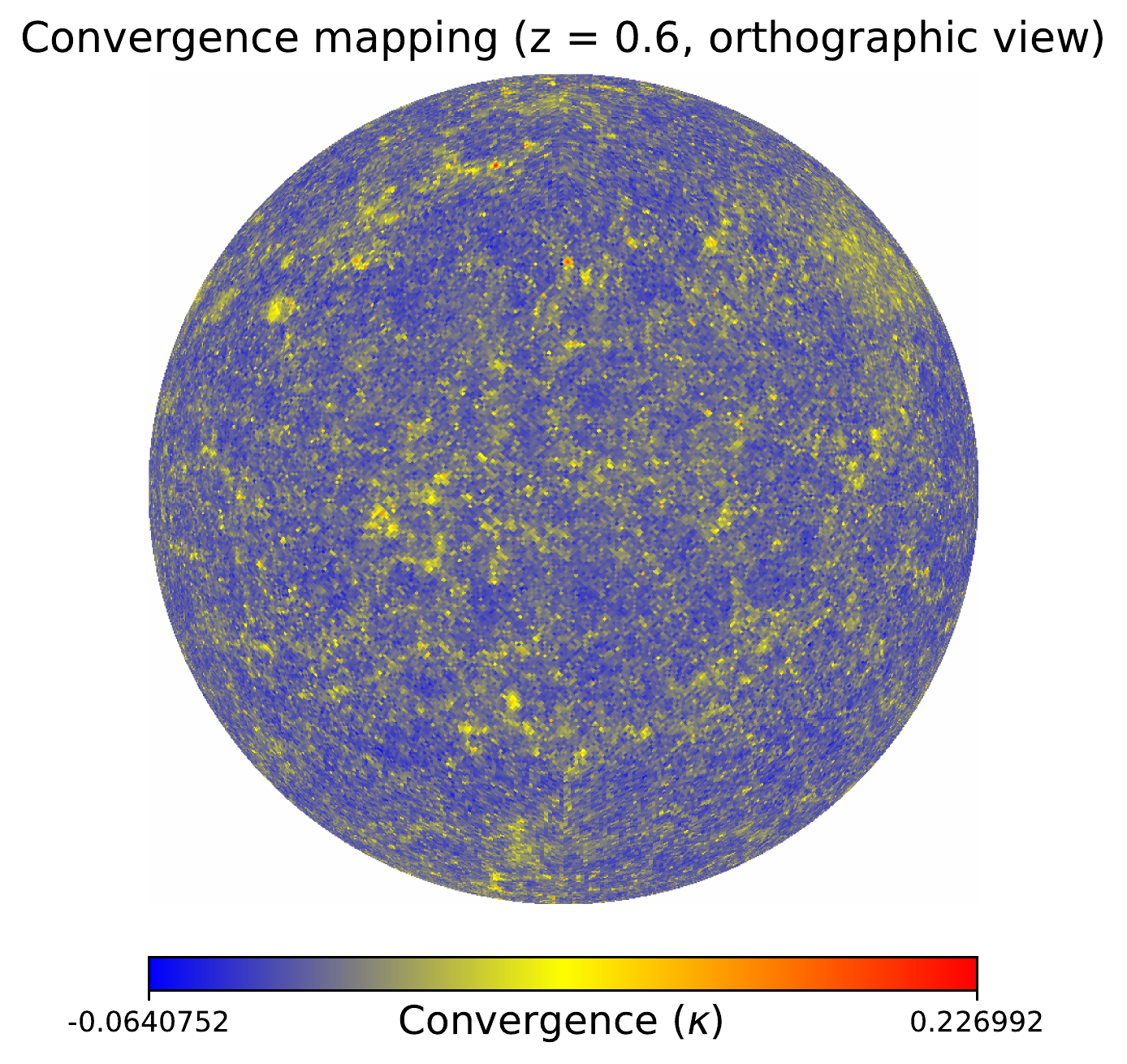}
            \end{subfigure}
    \caption{Maps of WL statistics at redshift $z$ = 0.6 (orthographic projection). The results are based on the computational approach described in Section \ref{sec:Mapping the WL statistics} and all of the three maps are taken from the same realization. Here the darker (lighter) color in the color bar indicates under-dense (over-dense) regions in the LSS. }
    \label{fig:MAPS}
\end{figure}

\begin{figure*}
    \centering
    \hspace*{-0.2in}
    \includegraphics[width=2\columnwidth]{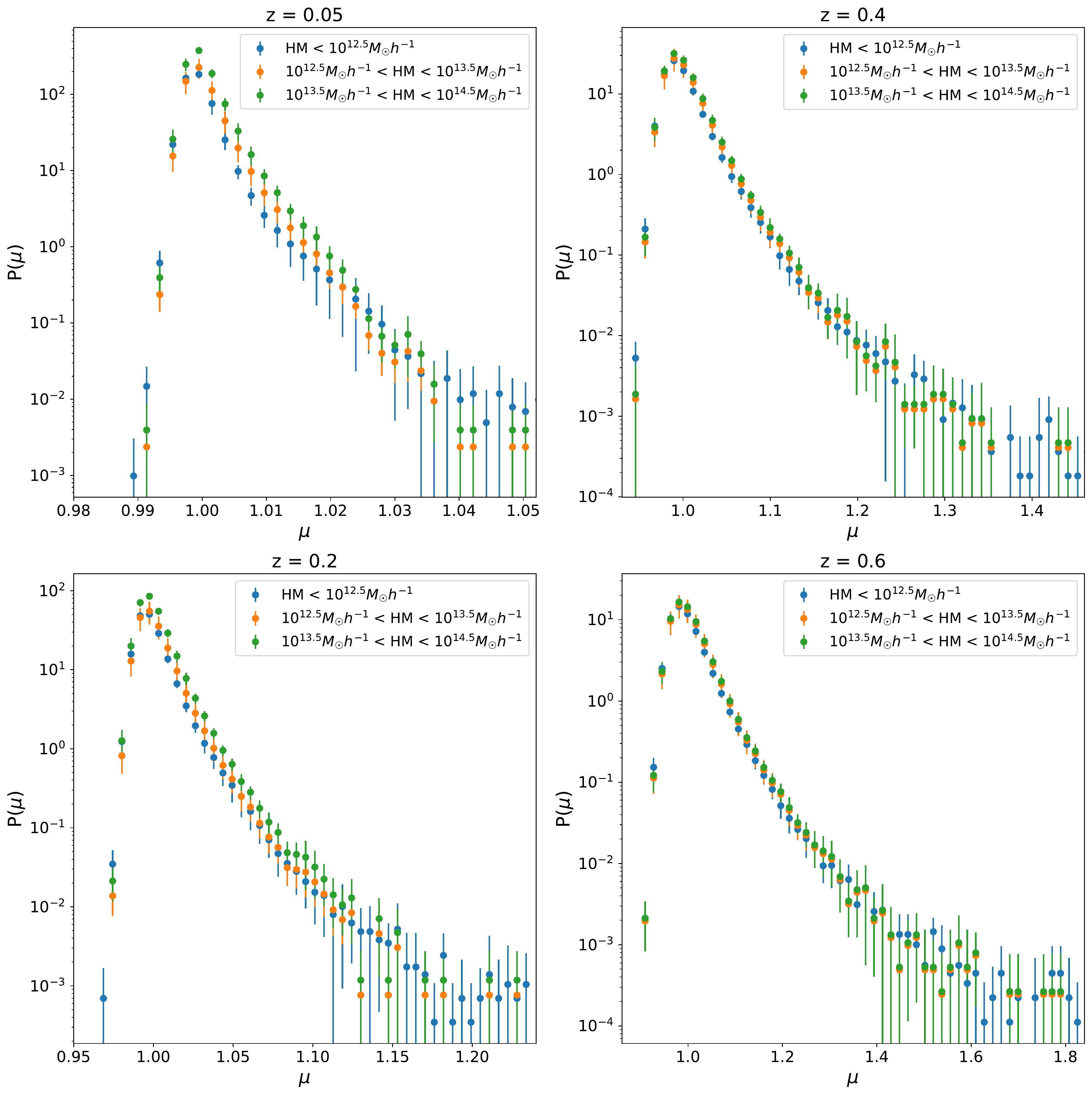}
    \vspace*{-3mm}
    \caption{Magnification PDFs as a function of redshift when observers were located in halos having different masses. Here the markers indicate mean PDFs and the error bars show 68\% data of magnification around the mean value at each mass range.}
    \label{fig:halo_mu_PDF}
\end{figure*}

\subsection{Mapping the WL statistics}
\label{sec:Mapping the WL statistics}

Mapping of WL properties provides information regarding the evolution of LSS of the universe \citep{Takahashi2012, Carbone2013, Lepori2020}. An example of such a map is presented in Fig. \ref{fig:MAPS}, which shows the orthographic projection of the WL maps for magnification, shear, and convergence at a redshift of 0.6.

The maps were obtained by applying the RBM. Here the bundles of geodesics propagated through a simulated universe based on the $N$-body code \texttt{gevolution}.
The bundles were propagated up to redshift of 0.6. Then, 
 Eq. \ref{mu_RBM} was used to calculate magnification, Eq. \ref{shear_RBM} to calculate cosmic shear, and Eq. \ref{mu_equation} to calculate the WL convergence. The patterns of magnification and convergence mapping look similar, whereas some filament type structures have been found in the map of cosmic shear. These results are similar to some previous studies \citep{Takahashi2017, Fabbian2018, Lepori2020}. This figure reflects the behaviour of LSS formation of the universe which is due to the impact of density perturbation. In each of the maps in Fig \ref{fig:MAPS}, darker regions indicate zones having low density, whereas lighter regions are the highly dense zones on the LSS of the universe. These maps are then analysed using one-point and two-point statistics.

\begin{figure*}
  \centering
  \hspace*{-0.2in}
  \includegraphics[width=2\columnwidth]{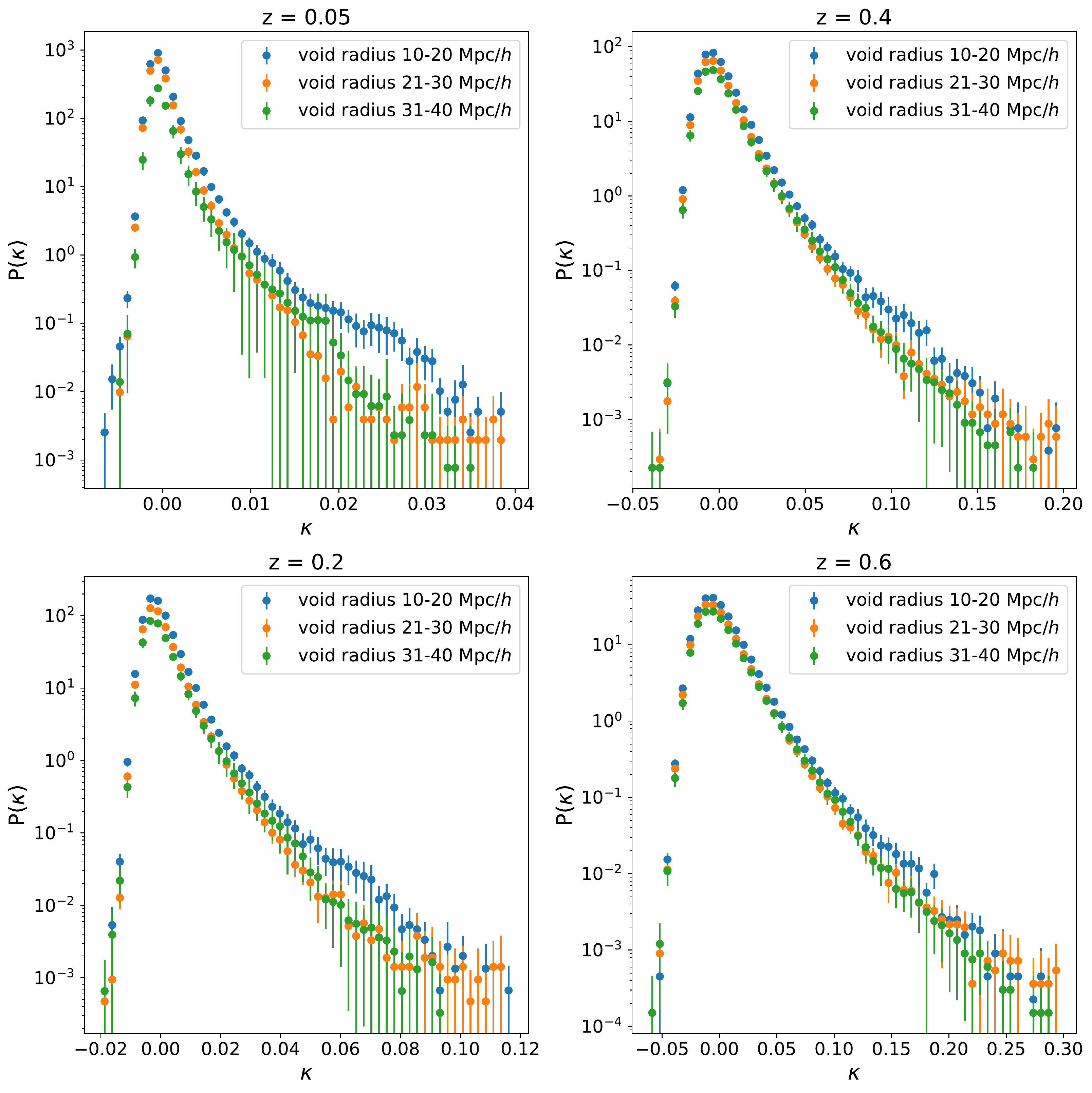}
  \vspace{-3mm}
  \caption{PDFs of convergence as a function of redshift when observers were located in voids having different radii. Here the markers indicate mean PDFs and the error bars having different colors show 68\% data around the mean value of convergence at each radius range.}
   \label{fig:convrg_PDF_void}
\end{figure*}

\section{SIMULATIONS}

\subsection{Simulating the LSS}

\label{sec:SIMULATION}

All the results that we will present in Section \ref{sec:RESULTS ANALYSIS} are from cosmological simulations, whereas the lensing potentials have been generated using the weak field metric from the code \texttt{gevolution} \citep{Adamek2016}. We develop a ray-tracing code that can solve null geodesic equations at any redshift and study the WL observables. We adopt a standard $\Lambda$CDM cosmology where the values of cosmological parameters are $\Omega_m$ = 0.312, $\Omega_{\Lambda}$ = 0.6879, $h$ = 0.67556, primordial
amplitude of scalar perturbations $A_s$ = 2.215 $\times$ $10^{-9}$ at the pivot scale $k_*$ = 0.05 $\mathrm{Mpc^{-1}}$, spectral index $n_s$ = 0.9619, and massless neutrinos with $N_\mathrm{eff}$ = 3.046, respectively. The parameter $h$ is used here only to convert the physical neutrino density to a dimensionless density parameter. In \texttt{gevolution}, the linear transfer function and the primordial power spectra were generated at an initial redshift of $z = 100$ with CLASS \citep{Blas2011}. We run around $150$ numerical simulations with different initial conditions, each having in total 256$^3$ dark matter particles in a simulation volume of $(320~\mathrm{Mpc}/h)^3$.

\subsection{Finding halos and voids}
\label{sec: Finding voids and clusters}
Before analysing any outcomes regarding the influence of our local universe, at first it's important to characterise potential locations of observers in terms of their place in the cosmic web. To do that, we find out the positions of all voids and halos on the LSS within our simulation volume. 

We employ two publicly available codes to identify the position of halos and voids from our simulation. Firstly, \texttt{ROCKSTAR}\footnote{\url{https://github.com/yt-project/rockstar}} \citep{Behroozi2013} is an algorithm that is based on an adaptive hierarchical refinement of friends-of-friends groups in six phase space dimensions and one time dimension.
This approach can be used to find the substructures as well as dark matter halos from cosmological simulations. Based on particle density, we use \texttt{ROCKSTAR} \citep{Behroozi2013} in this work to identify the halo positions, masses, velocities, and other halo properties from our simulation data. We use the particle snapshot at redshift $z$ = 0 from \texttt{gevolution}, for the same simulation setting as described in \ref{sec:SIMULATION}. We find in total 30,321 halos using \texttt{ROCKSTAR} and then cluster all of these halos according to their mass ranges. According to the mass information from \texttt{ROCKSTAR}, we find total 17,613 halos when the halo mass is smaller than $10^{12.5}$ $M_\odot$ $h^{-1}$, in total 11,353 halos when $10^{12.5}$ $M_\odot$ $h^{-1}$ < halo mass < $10^{13.5}$ $M_\odot$ $h^{-1}$, and total 1,300 number of halos have been found where $10^{13.5}$ $M_\odot$ $h^{-1}$ < halo mass < $10^{14.5}$ $M_\odot$ $h^{-1}$. 

To identify voids, we adopt the void finder \texttt{Pylians}\footnote{\url{https://github.com/franciscovillaescusa/Pylians}}.
This can be used to analyse the spherical voids from the results of both $N$-body and hydrodynamic simulations. Besides the identification of voids, \texttt{Pylians} can also be used to compute power spectra, bispectra, correlation functions, and density fields. But in this study, we pay attention only to find the positions and radii of the voids because the observers' location is the foremost priority here. From the particle's snapshot at redshift $z$ = 0, we identify the positions and radii of all voids, when the simulation setting was the same as we mentioned in Sec. \ref{sec:SIMULATION}. We find in total 5,261 voids by \texttt{Pylians} and then categorize them according to their radii ranges (e.g., range 1: 10 - 20 $\mathrm{Mpc}/h$; range 2: 21 - 30 $\mathrm{Mpc}/h$; range 3: 31 - 40 $\mathrm{Mpc}/h$). There are around 2,559 voids in range 1, 342 voids in range 2, and around 126 voids have been found within range 3.

\begin{figure}
  \centering
  \includegraphics[width=\columnwidth]{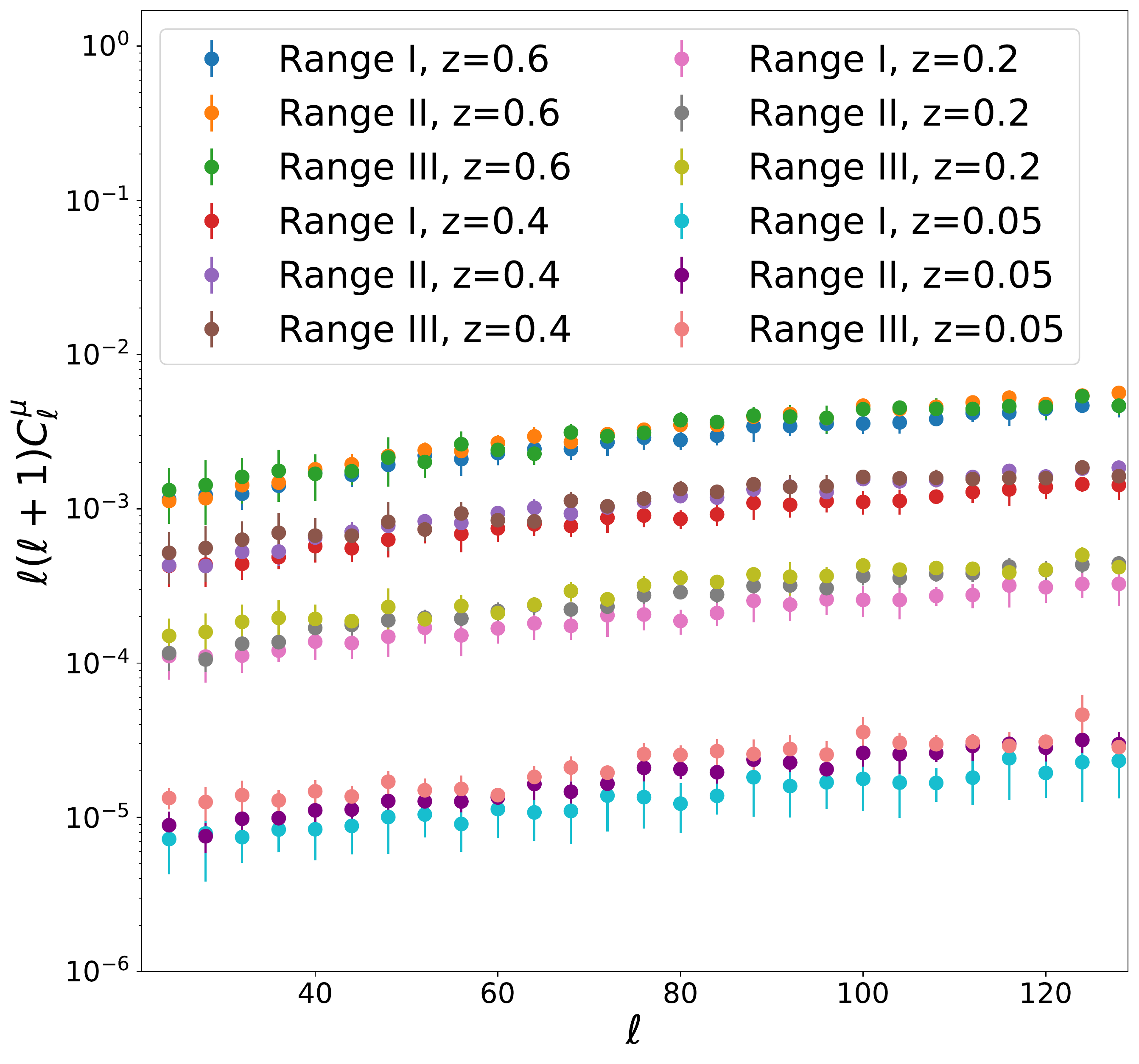}
  \caption{Angular power spectra of magnification as a function of redshift when observers were located in halos having different masses (Range I: halo mass < $10^{12.5}$ $M_\odot$ $h^{-1}$; Range II: $10^{12.5}$ $M_\odot$ $h^{-1}$ < halo mass < $10^{13.5}$ $M_\odot$ $h^{-1}$; Range III: $10^{13.5}$ $M_\odot$ $h^{-1}$ < halo mass < $10^{14.5}$ $M_\odot$ $h^{-1}$). Here markers indicate mean values and error bars having different colors show 68\% data around the mean angular power spectrum  of magnification for different redshifts at each mass range.}
   \label{fig:halo_mu_power_spectra}
\end{figure}

\section{The convergence angular power spectrum}

Here we consider the convergence angular power spectrum. This is obtained by comparing the expected theoretical angular power spectrum of the convergence with the actual angular power spectra inferred from ray-tracing within our numerical simulations. 

\subsection{Theoretical angular power spectrum of the convergence}
\label{sec: Theoretically calculate convergence angular power spectra}

The theoretical angular power spectrum is calculated by implementing the Limber approximation in the WL limit \citep{Limber1953, Krause&Hirata2010, Bartelmann&Maturi2016, Lemos2017, Kilbinger2017, Wei2018}, which makes an approximation of a flat sky and neglects the curvature of the sky. This approach is suitable for any survey areas with an extent less than 10 degrees \citep{Kilbinger2017}. Within this approximation the convergence is first related to the gravitational lensing potential $\phi$ and can be written in the form of the weighted integral of the overdensity $\delta$ along the line of sight as 
\begin{equation}
\begin{aligned} 
\kappa(\theta, \chi) = \frac{3 H_0^2 \Omega_m}{2 c^2} \int^0_\chi d\chi^{'}   \frac{r (\chi - \chi^{'}) r(\chi^{'})}{r(\chi)}  \frac{\delta(r(\chi^{'}) \theta, \chi^{'})}{a(\chi^{'})},
\end{aligned} 
\label{convrg}
\end{equation}
\noindent where $\theta$ is the observed angular position, $H_0$ is the Hubble constant, $\Omega_m$ is the matter density, $\chi$ is the comoving distance, $r(\chi)$ is the comoving angular diameter distance, and $a(\chi^{'})$ is the scale factor at $\chi^{'}$. Then the two-point correlation function in Fourier space can be considered as the power spectrum of convergence $C^{\kappa}(\ell)$, in the flat-sky limit. 
\begin{equation}
  \langle\tilde{\kappa}(\bm{\ell}) \tilde{\kappa}^{*}(\bm{\ell}')\rangle=(2\pi)^2 \, \delta_{\rm D}(\bm{\ell}-\bm{\ell}') \, C^{\kappa}(\ell),
\end{equation}
\noindent where $\delta_{\rm D}(\bm{\ell})$ is  the Dirac  delta function. Finally, the angular power spectra of convergence in the Limber approximation is
\begin{equation}
  C^{\kappa}(\ell) = \int^{\chi_{\rm H}}_{0} {\rm d}\chi \frac{W(\chi)^{2}}{r(\chi)^{2}} P_{\delta}\left(k=\frac{\ell}{r(\chi)}, \chi\right),
\end{equation}
\noindent where  $P_{\delta}(k,\chi)$ is the three dimensional matter power spectrum at the given comoving distance $\chi$  and $\chi_{\rm H}$ is the comoving horizon distance. Here the weight function $W(\chi)$ can be expressed as
\begin{equation}
  W(\chi) = \frac{3H_{0}^{2}\Omega_{{\rm m}}}{2c^2}\frac{r(\chi_{\rm H}-\chi)r(\chi)}{r(\chi_{\rm H})} \frac{1}{a(\chi)}.
\end{equation}
\noindent Here we use the publicly available code \texttt{CAMB}\footnote{\url{https://github.com/cmbant/CAMB}} (Code for Anisotropies in the Microwave Background) to calculate the theoretical power spectrum of the convergence for the $\Lambda$CDM model \citep{Lewis2000}. 
Due to the Limber approximation being valid for small angle approximation in this study we only take the angular power spectrum of convergence for the multipole $\ell > 10$  \citep{Lemos2017}.

\subsection{Inferring the angular power spectrum of the convergence from simulations} \label{sec: Numerically calculate convergence angular power spectra}

 \cite{Peebles1973} proposed a technique to estimate the angular power spectra $C_{\ell}$, where projected density field onto the plane of the celestial sphere was decomposed into spherical harmonics. If there is a bandlimited function $f$ on the sphere then it can be expanded in spherical harmonics, $Y_{\ell m}$, as
\begin{equation}
f(\zeta) = \sum_{\ell = 0}^{\ell_{max}} \sum_m a_{\ell m} \, Y_{\ell m}(\zeta).
\end{equation}
\noindent Here $a_{\ell m}$ is the spherical harmonic coefficients, ${{\zeta}}$ denotes a unit vector pointing at polar angle $\theta\in[0,\pi]$, and azimuth $\phi\in[0,2\pi]$. This spherical harmonic coefficient $\hat{a}_{\ell m}$ can be used to compute the angular power spectrum $\hat{C}_\ell$ as 
\begin{equation}
\hat{C}_\ell = \frac{1}{2\ell + 1} \sum_m |\hat{a}_{\ell m}|^2.
\label{power_spectra}
\end{equation}

\noindent This is a direct method of inferring the angular power spectrum. In our analysis that above mentioned function $f$ is the WL convergence. Thus, in order to infer the angular power spectrum from our ray-tracing simulations we first generate WL maps (cf. Fig. \ref{fig:MAPS}). In our study, each map is generated by sending ray bundles in separate directions, corresponding to separate pixels on the HEALPix sphere \citep{1999astro.ph..5275G}. 
The Euler-Rodrigues formula has been used here to initially align each bundle with its corresponding normal vector. For each bundle we then infer its magnification and shear, see Sec. \ref{sec:magnification} and \ref{sec:shear}. This is then used to calculate the WL convergence by the means of eq. (\ref{mu_equation}). The convergence maps are then analysed using the code \texttt{healpy}\footnote{{\url{https://github.com/healpy/healpy}}}, which is a python code that implements the HEALPix scheme \citep{2005ApJ...622..759G}.
Finally, we compute the angular power spectra using the  function \texttt{healpy-anafast}.

\section{RESULTS AND ANALYSIS}
\label{sec:RESULTS ANALYSIS}

\subsection{Local environment influence on one-point WL statistics}

We present here results from the one-point PDFs for various WL properties. All of the results discussed in this subsection contain descriptive information about the local cosmological environments (i.e. the effect on WL observables due to the positions of the observer within the LSS of the universe).

Using our ray-tracing algorithm we generate the PDFs from the WL statistics. After identifying the cosmic structures from the snapshot at today's redshift, we categorize the halos according to their mass ranges and voids according to their radii ranges. We set the observers in different locations, here in halos having different masses and in voids having different radii, then solve around 0.4 million geodesics by projecting ray bundles starting from redshift $z$ = 0 backwards in time using our ray-tracing algorithm and calculate the PDFs for each realization. Finally, we compute the mean PDFs for each range of halo masses and void radii, then generate the error bars from 68\% confidence interval (CI) around every mean PDFs.

Figure \ref{fig:halo_mu_PDF} shows the WL magnification PDFs of sources at different redshifts. Error bars indicate 68\%  data around mean for different halo mass ranges and the scaling of the x-axis for the magnification is different for each redshift. It is clearly seen that for more distant sources, the width of the distributions becomes broader and the peak moves toward demagnification (where $\mu < 1$). These aspects are compatible with some previous studies \citep{Wambsganssr1997, Wang2002, Hilbert2007, Takahashi2011,2012JCAP...05..003B}. We find higher magnification for observers in massive halo regions as compared to less massive halo regions. We then place the observer in the underdense region.
Fig \ref{fig:convrg_PDF_void} shows how the WL observables for sources at different redshifts and accordingly to the radius of a void where the observer resides in. The results are consistent with the one for observers residing in overdense regions: the denser the environment the higher the amplitude of the PDFs.

\subsection{Local environment influence on two-point WL statistics}

\begin{figure}
  \centering
  \includegraphics[width=\columnwidth]{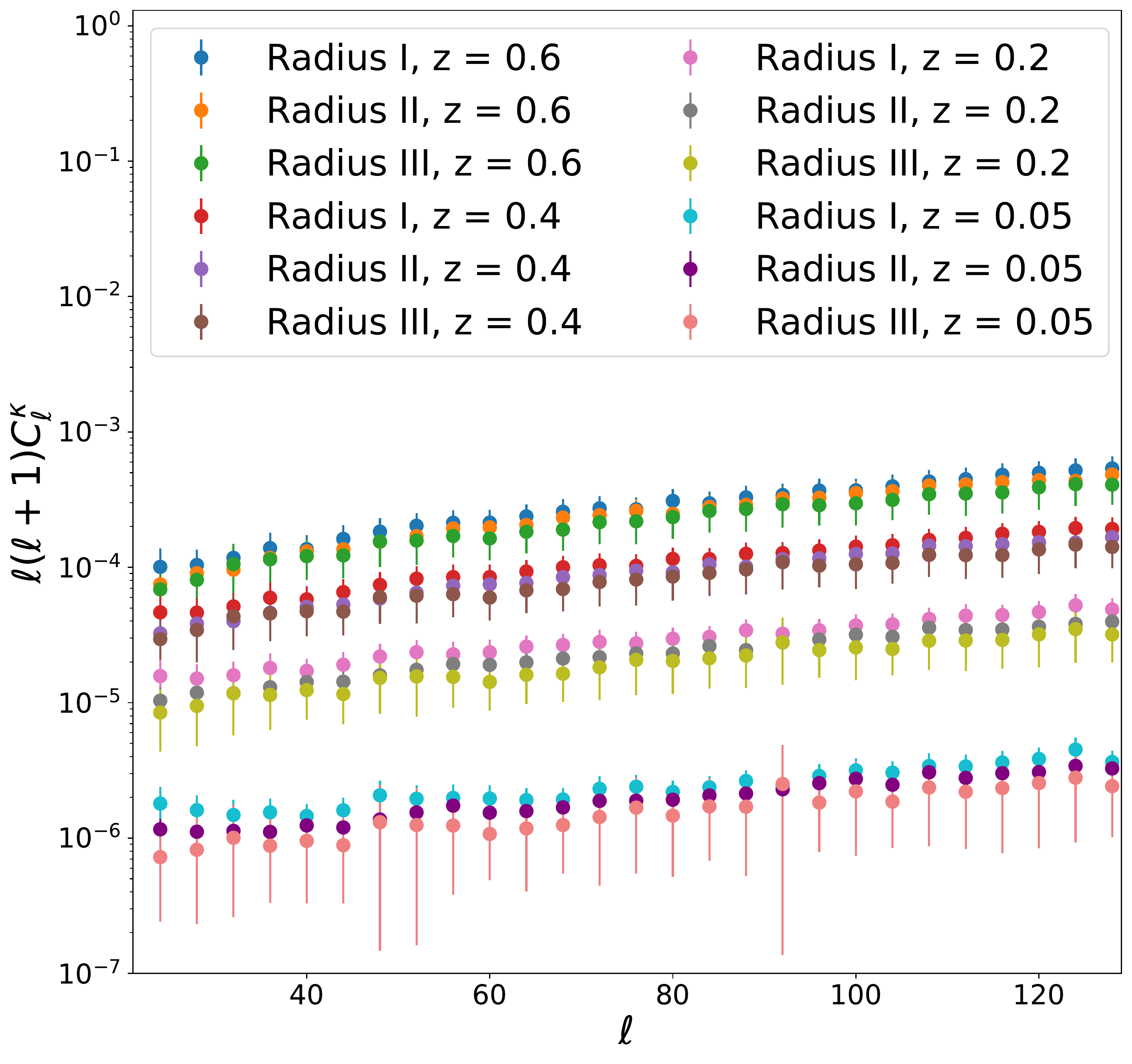}
  \caption{Angular power spectra of convergence as a function of redshift when observers were located in voids having different radii (Radius I: 10-20 Mpc/$h$; Radius II: 21-30 Mpc/$h$; Radius III: 31-40 Mpc/$h$). Here markers indicate mean values and error bars having different colors show 68\% data around the mean angular power spectrum  of convergence for different redshifts at each radius range.}
   \label{fig:convrg_spectra_void}
\end{figure}

In this section, we present results regarding the effect of the local cosmological environment on the angular power spectra. Here we also place the observers in halos having different masses and in voids having different radii.

Fig. \ref{fig:halo_mu_power_spectra} shows how the angular power spectrum of magnification changes for sources at different redshifts and accordingly to the masses of the halos where the observers reside in. 
The analysis of the angular power spectrum of voids having different radii is presented in Fig. \ref{fig:convrg_spectra_void}. Both of these figures show 
the same phenomena as we have seen from the analysis of PDFs for halos having different masses and voids having different radii, i.e. the amplitude of the power spectrum depends on how dense the local cosmological environment is.
As before, the effect decreases with redshift, which is understandable as signal becomes more dominated by the underlying matter distribution along the line of sight, rather than the local environment.

\subsection{The effect on the estimation of cosmological parameters}

Our goal is to investigate the effect of the local cosmological environment on WL observables. 
In particular we ask what is the minimal redshift beyond which the effects due to the local environment are negligible. To do that, we focus on constructing likelihood contours and provide constraints on cosmological parameters from the WL statistics of simulation data.
 
The first step is to measure the angular power spectra of the WL convergence from the data of numerical simulation. Firstly, the observer is placed in a random halo of mass $M = 10^{12.5} M_\odot$. Then, we calculate the WL convergence angular power spectra at different redshifts. We then generate model data from the theoretical predictions, see Sec. \ref{sec: Theoretically calculate convergence angular power spectra}. 
Finally, we use the Markov chain Monte Carlo (MCMC) method to estimate the posteriors of the cosmological parameters ($\Omega_m$, $H_0$) for a given parameter space and generate the likelihood contours. 
The  MCMC analysis is done using the code \texttt{emcee}\footnote{\url{https://github.com/dfm/emcee}}. In this study, we use uniform priors in the parameter ranges $\Omega_m: [0.1, 1.2]$ \& $H_0:[30, 100]$ and the posterior probability outside this prior range is 0. 

\begin{figure}
  \centering
  \includegraphics[width=\columnwidth]{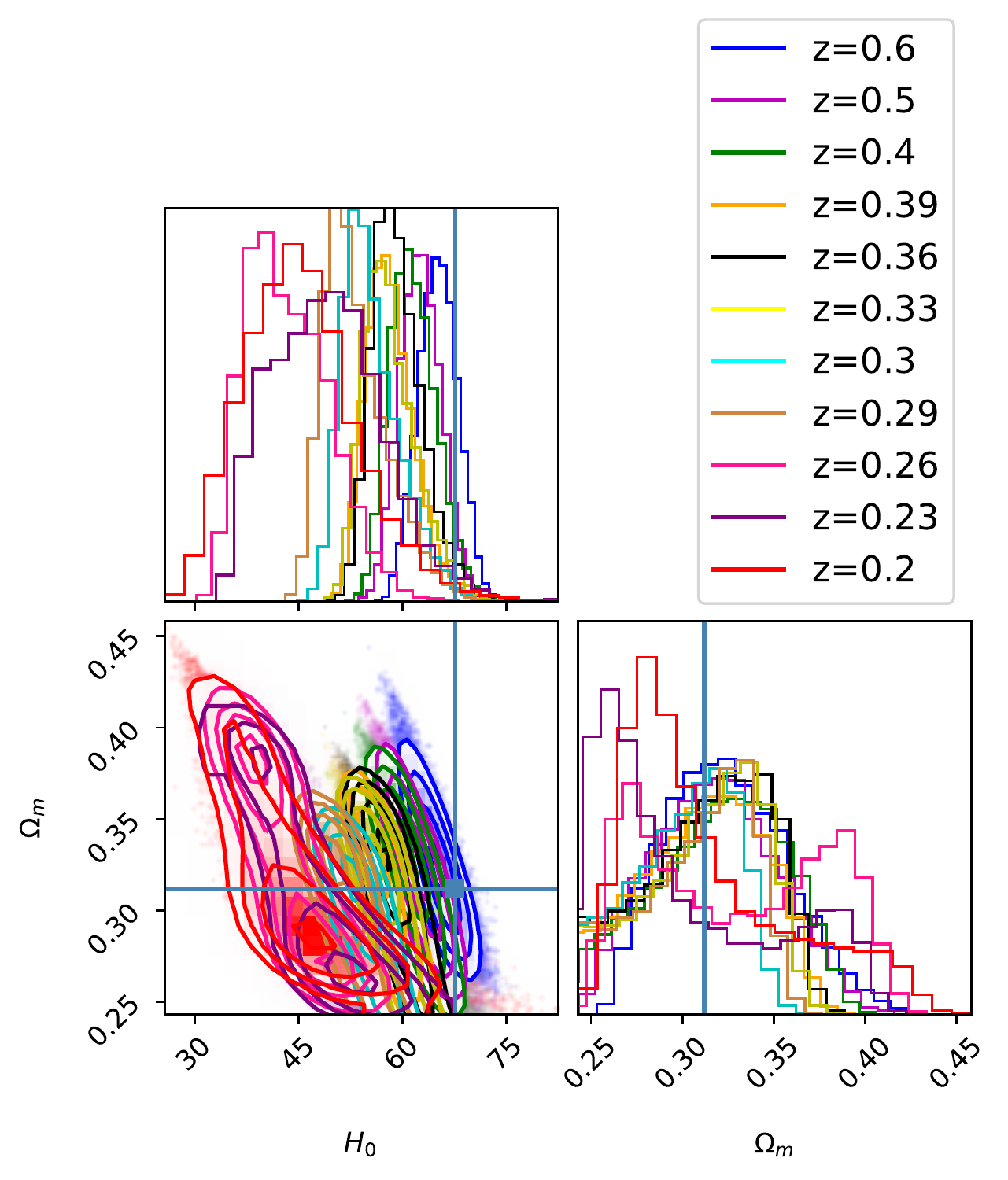}
  \caption{Expected constraints on cosmological parameters as a function of redshift of the data used to constraints these parameters. 
The constraints are based on mock data from simulations with $H_0 = 67.556 \, {\rm km} \, {\rm s}^{-1}\, {\rm Mpc}^{-1}$ and $\Omega_m = 0.312$. The mock data consists of the angular power spectrum of WL convergence of 49,152 sources.
The angular power spectrum is then analysed against a theoretical convergence power spectrum that is sensitive to $H_0$ and $\Omega_m$.}
   \label{fig:constraints}
\end{figure}

\begin{figure}
  \centering
  \noindent
  \includegraphics[width=\columnwidth]{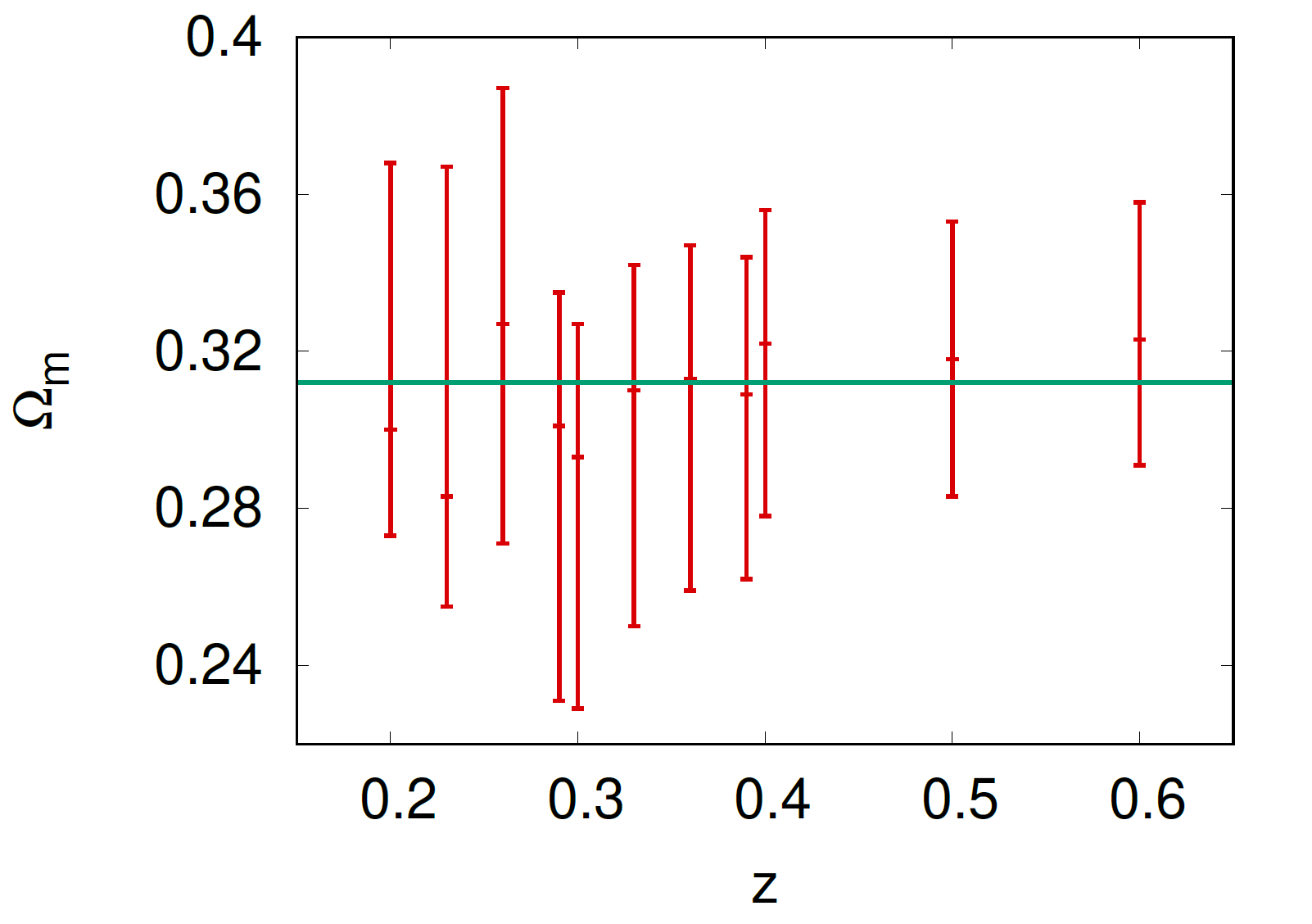}
  \caption{Constraints on the parameter $\Omega_m$ as a function of source redshift. The constraints are based on the angular power spectrum of WL convergence of 49,152 sources, simulated using $\Omega_m = 0.312$. 
The constraints are similar to those presented in Fig. \ref{fig:constraints} but marginalised over $H_0$. } 
   \label{fig:omega}
\end{figure}

\begin{figure}
  \centering
  \noindent
  \includegraphics[width=\columnwidth]{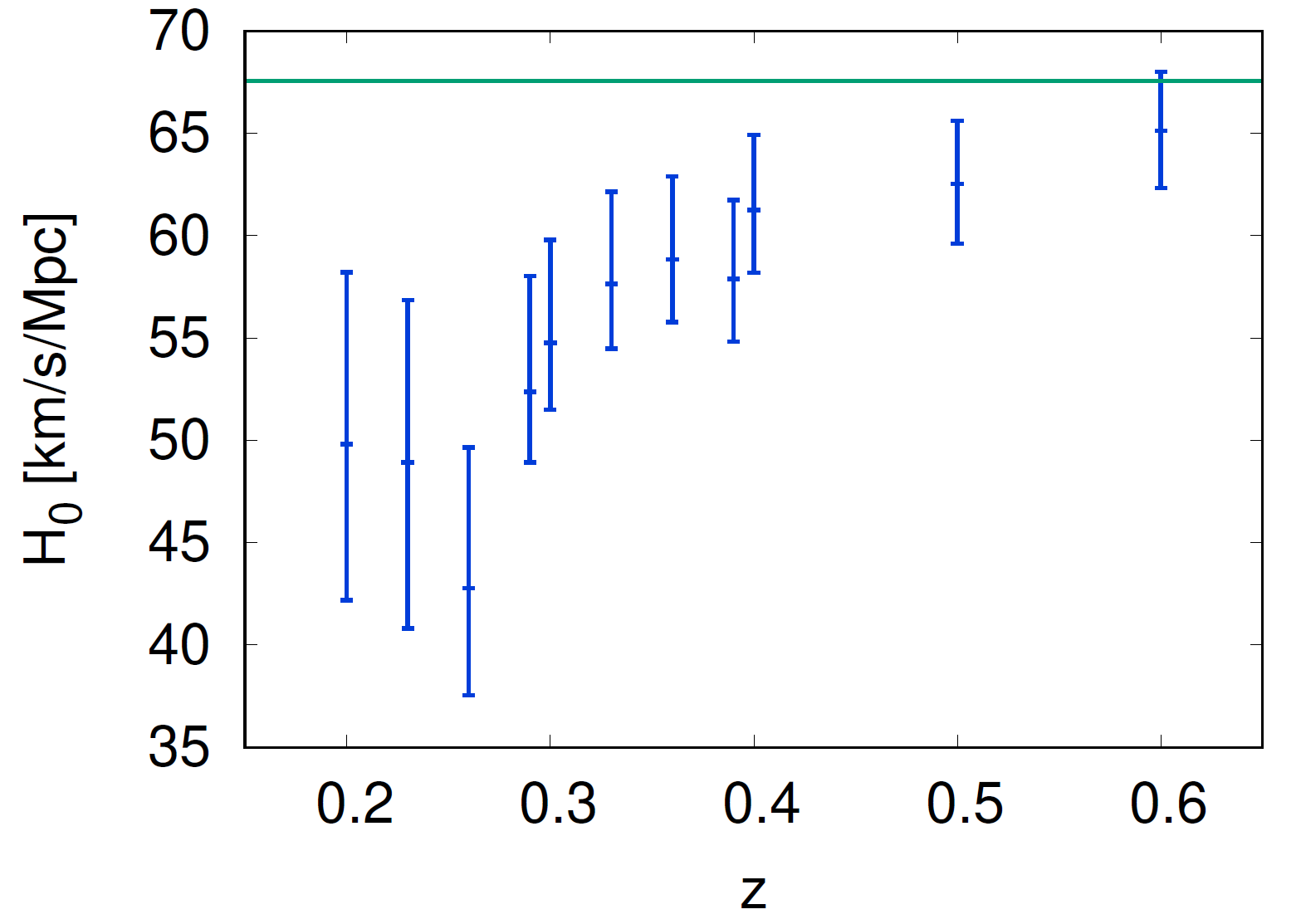}
  \caption{Constraints on the parameter $H_0$ as a function of source redshift. The constraints are based on the angular power spectrum of WL convergence of 49,152 sources, simulated using $H_0 = 67.556 \, {\rm km} \, {\rm s}^{-1}\, {\rm Mpc}^{-1}$. The constraints are similar to those presented in Fig. \ref{fig:constraints} but marginalised over $\Omega_m$.}
   \label{fig:hubble}
\end{figure}

The results are presented in Fig \ref{fig:constraints}, which shows the corner plot for the two-dimensional posterior probability distributions of the cosmological parameters as a function of redshift. In this figure, the blue horizontal line shows our true cosmology, $H_0 = 67.556$ $\mathrm{km/s/Mpc}$ \& $\Omega_m = 0.312$, and the posterior distributions indicate how the WL properties are influenced by the local universe at different redshifts. This plot has been generated from the same locations of the observers but at different redshifts of the sources. 
It should be emphasised that for each constraints we have the same amount of sources. Unlike in the case of realistic WL surveys where the number of galaxies increases with redshift, here regardless the redshift we have the same amount of bundles, i.e. 49,152 ray bundles. 
Figures \ref{fig:omega} and \ref{fig:hubble} show how constraints on the cosmological parameters $\Omega_m$ and $H_0$ respectively, change as a function of redshift. 

These results consistent with those reported in \citet{Reischke2019} that local cosmological environment affects the WL observables, the only difference between the analysis of the paper and the results presented in \citet{Reischke2019} is that here we confirm these using the relativistic $N$-body simulations and by solving geodesic equations.

\section{CONCLUSIONS}
\label{sec:CONCLUSIONS}

The local cosmic structure or our observable universe includes cosmic voids, galaxy, clusters of galaxy, filaments, and many other structures also. All of these shapes contribute to the cosmic web and solving geodesic equations is a proper way to study the LSS as well as gravitational lensing for different lens approximations \citep{Kasai1990, Killedar2012, Mood2013, Lepori2020}. The view of the LSS is not identical and the local structures are formed there due to the variation of particle mass distribution or density fluctuations. That's why all of the images of galaxies, that are situated far away from us, may not be looked like similar to all of the observers who lay down in different zones on the LSS. The physics of voids and halos depending on their mass ranges, radius ranges, and density distributions have been analysed through many works already  \citep{Amendola1999, Bolejko2013, Mood2013, Davies2018, Fong2018}. 

This paper has numerically probed the influence of the local environment on WL statistics depending on where the observer is located. To do this study-we have constructed a ray-tracing algorithm that is a well-organised code and easy to implement with any simulations to study the LSS of the universe.  We have also studied the posterior distributions of the cosmological parameters to see how the local environment's effect changes as a function of redshift and beyond which redshift such effects on the data from any cosmological observations is negligible. We have constructed a ray-tracing code here, that uses the gravitational potentials from a relativistic $N$-body simulation, to solve the geodesic equations by varying the observers' location in different void and cluster regions. To project the light rays in different directions in the sky, we have implemented RBM where we project light rays as bundles instead of a single light ray. We have run a number of numerical simulations and studied the maps, PDFs, and angular power spectrum of various WL characteristics. Further, we have adopted the Bayesian statistics approach to analyse the constraints on cosmological parameters as a function of redshift. The distributions are very sensitive at low redshifts, the curves are getting closer to the fiducial cosmology as the light travels towards high redshifts.

According to the cosmological perturbation theory, today's LSS was formed due to the growth of density fluctuations from the early universe. Using perturbation theory, \cite{DiDio2014} theoretically compared the standard Fourier power spectra with the redshift dependent angular power spectra of galaxy number counts and studied the sensitivity of future Euclid-like \citep{Amendola2018} galaxy surveys. The effect of neglecting lensing magnification in the future galaxy surveys like SKA\footnote{\url{https://www.skatelescope.org}} and Euclid \citep{Amendola2018} was investigated by \cite{Villa2018} considering three cosmological models, one is standard $\Lambda$CDM and the others are two extensions: massive neutrinos and the modifications of GR. Again, the amplitude of relativistic contributions to the WL convergence power spectra was estimated by \cite{Andrianomena2014} where the gravitational wave and frame-dragging rotation were taken into consideration as backgrounds. Unlike those previously mentioned studies the goal of this present work is different and here we have numerically analysed that the observer will observe non-identical scenarios when they are located in voids having different radii and halos having different masses. We have also focused on the cosmological parameter estimation and studied numerically how the posterior distributions are changing with the redhsift of the sources. The results of this analysis confirms that local cosmological environment has the effect on low redshift data, and one needs to use sources beyond some minimal redshift beyond which the effect is negligible.

This study concludes that when inferring cosmological parameters from observations data one need to account for effect of the local cosmological environments. This study suggests that the minimal redshift of WL surveys from which the constraints on $\Omega_m$ can be meaningful is around $z > 0.2$, whereas for the parameter $H_0$ we require higher redshift, i.e. $z> 0.6$. These results were obtained based on 49,152 ray bundles, and thus the relevance for realistic WL surveys especially depends on the number of data remains to be investigated. As the goal of all future WL surveys is to study the LSS with very good precision, this study can contribute to produce more meaningful results from the data of any future cosmological observations. This paper contains the results from the analysis of the PDFs and angular power spectrum of WL statistics. It would be also interesting to investigate the impact of our local structures on the WL bispectrum and also search for the redshift from which there are no such effects on the data from any WL surveys, we left these for our future study.

\section{ACKNOWLEDGEMENTS}
\label{sec:ACKNOWLEDGEMENTS}
The authors would like to acknowledge Artemis, HPC support at the University of Sydney, for providing computational resources that has contribution to produce the results reported in this paper. SAE thanks Florian List for his comments on this paper. SAE is supported by the Commonwealth Government funded Research Training Program (RTP) Stipend Scholarship. SAE is grateful to her child Raheen Hossen Ilhaan for his unaccountable support from mother's womb while preparing this paper. KB acknowledges support from the Australian Research Council through the Future Fellowship FT140101270. This work has made use of \texttt{numpy} \citep{Harris2020}, \texttt{h5py}\footnote{\url{https://www.h5py.org/}}, \texttt{matplotlib} \citep{Hunter2007}, \texttt{mpi4py} \citep{Dalcin2005}, and \texttt{eqtools}\footnote{\url{https://eqtools.readthedocs.io/en/latest/}}.

\section{DATA AVAILABILITY}
\label{sec:DATA AVAILABILITY}
The data generated as part of this article will be shared on a reasonable request to the corresponding author.

\bsp	
\label{lastpage}
\end{document}